\documentclass[12pt,,letterpaper]{JHEP3}
\usepackage{graphics}
\usepackage{epsfig}
\usepackage{slashed}

\title{Holographic fermions in charged Lifshitz theory}
\author{Li Qing Fang ${}^{1}~$, Xian-Hui Ge ${}^{1}~$, Xiao-Mei Kuang${}^{2}~$ \\
${}^{1}$Department of Physics, Shanghai University, 200444 Shanghai,
China\\
${}^{2}$INPAC and Department of Physics, Shanghai Jiaotong University, 200240 Shanghai, China\\
\email{flqthunder@163.com},~~\email{gexh@shu.edu.cn},~~\email{xmeikuang@gmail.com} }

\abstract{We investigate the properties of holographic fermions in charged Lifshitz black
holes at finite temperature through the AdS/CFT correspondence. In the charged Lifshitz
background with the dynamical exponent $z=2$, we find that the dispersion
relation is linear. The scaling behavior of the imaginal part of the Green function
relative to $k_{\perp}=k-k_F$ is also discussed. We find, although the system has linear dispersion relation and quadratic quasi-particle width, it does not satisfy Luttinger's theorem. We also find that the variation of the scaling parameters $\alpha$ and $\beta$ is small as the charge $q$ varies. Furthermore, we also discuss the effect of the dynamical exponent $z$ by considering the cases $z=4$ and $z=6$ and show that  $ImG_{ii}$ become smooth when the dynamical exponent $z$ increases.}

\begin{document}

\section{Introduction}
Recently, there has been an increased interests in studying  strongly interacting fermionic  systems by a gravity dual using the holographic Anti-de Sitter/conformal field theory correspondence (AdS/CFT) or Anti-de Sitter/condensed matter theory (AdS/CMT) correspondence \cite{adscft,gkp,w}. In \cite{f1,f2,f3,f4}, it was shown that holographic duality can describe Fermi and non-Fermi liquids.
In general, the non-Fermi liquid has a sharp Fermi surface but its gapless charged excitations  differ significantly from those predicted from Landau's Fermi liquid theory. Up to now, a proper theoretical framework characterizing non-Fermi liquids is lacking. Fortunately, the AdS/CMT correspondence provides a large theoretical playground of non-Fermi liquids and give a handle on a difficult strong coupling problem.

In many condensed matter systems, one can find phase transitions governed by fixed points with Lifshitz dynamical scaling $t\rightarrow \lambda^z t$, $x\rightarrow \lambda x$. Most interesting examples arising in the condensed matter systems correspond to the anisotropic scale invariance with $z=2$ that describes multicritical points in certain magnetic materials and liquid crystals, although when $z=1$  the scaling describes scale invariance which arises in the AdS/CFT correspondence. In fact, there are many non-relativistic $z\neq 1$ fixed points in condensed matter system including ulta-cold atom mixtures, Fermions at unitarity  and Bi-layer graphene. The gravity dual of such non-relativistic field theories was first realized  by Kachru, Liu, and Mulligan in \cite{lif0}.

In this paper, we study the fermionic systems numerically in charged Lifshitz theory at finite temperatures. Black holes in asymptotically Lifshitz spacetime provide a window onto
finite temperature effects in strongly coupled Lifshitz models. Black hole solutions and their related properties in Lifshitz spacetime  were studied by several authors \cite{lif1,lif2,lif3,lif4,lif5,lif6,lif7,lif8,lif90}.
Charged  Lifshitz black hole solutions were constructed in \cite{lif9} by adding a second Maxwell field $\mathcal {F}_{(2)}$ and a scalar field $\psi$
which was charged under the new gauge field but neutral under the original gauge fields $ {F}_{(2)}$ and $ {H}_{(2)}$. Later, this was generalized  to an arbitrary $(d+2)$ dimensional spacetime in \cite{DWP}. A much more general class of solutions of charged Lifshitz black holes was discussed numerically in \cite{DMP}.
In the very recent paper \cite{sto}, the retarded single-particle Green's function is derived in the Lifshitz background for a fermion coupled to the conformal field theory at vanishing spatial momenta, temperature and in the absence of the chemical potential.
The characters of non-relativistic fermionic retarded Green's function in Lifshitz geometry with critical exponent $z=2$ \cite{lif5} were studied in \cite{AMM}. In this paper, we mainly focus on the charged Lifshitz black hole and its gravity duals given in \cite{DWP} with dynamical exponents $z=2$ (i.e. three-dimensional black holes), $z=4$ (i.e. four-dimensional black holes) and $z=6$ ( i.e. five-dimensional black holes). We find that the charge $q$ and the exponents $z$ can affect the properties of the holographic fermions in different ways.

The rest of the paper is organized as follows: We begin by giving a brief review of the
charged Lifshitz black hole solutions. In section 3, we investigate the Dirac equation as well
as the boundary conditions in charged Lifshitz black hole backgrounds. By solving the Dirac
equation numerically, we discuss the properties of the holographic fermions, such as the values of the Fermi momenta $k_F$ and scaling
behaviors with different charge $q$, in $z=2$ charged Lifshitz black hole background in the first part of section 4. After that, we turn to $z=4$ and $z=6$ cases. We present the conclusion in the last section.

\section{Charged Lifshitz Black Holes}

In this section, we briefly review on the charged Lifshitz black hole solutions. We start with
the $d+2$ dimensional spacetime with the action
\begin{equation}\label{a}
S=\frac{1}{16 \pi G_{d+2}}\int d^{d+2}x \sqrt{-g} (R-2\Lambda-\frac{1}{4}F_{\mu\nu}F^{\mu\nu}-\frac{1}{2}m_{vector}^{2} A_\mu A^\mu-\frac{1}{4}\mathcal{F}_{\mu\nu}\mathcal{F}^{\mu\nu}),
\end{equation}
where $F_{\mu\nu}$ and $\mathcal{F}_{\mu\nu}$ are gauge fields strength. Here $\mathcal{F}_{\mu\nu}$ is
introduced to obtain charged black hole solution under the massive ``gauge field" $F_{\mu\nu}$
in this spacetime. Meanwhile $F_{\mu\nu}$ plays a key role in modifying the asymptotic geometry from AdS to
Lifshitz. Note that the massive vector field $A_{\mu}$ cannot directly be associated with a global $U(1)$ gauge field because there are no such
gauge degrees of freedom, though its kinetic terms look like a Maxwell term \cite{lif9,DWP}. Therefore, we will utilize the $\mathcal{F}_{\mu\nu}$ field  with global $U(1)$ symmetry in which a non-zero charge density can be introduced in what follows.

Then based on the action (\ref{a}), the black hole solution has the form\cite{lif9,DWP}
\begin{eqnarray}\label{blackhole}
ds^2=L^2[-r^{2z}f(r)dt^2+\frac{1}{r^2}\frac{dr^2}{f(r)}+r^2\sum_{i=1}^{d}dx_{i}^{2}],\nonumber\\
f(r)=1-\frac{r_{+}^{z}}{r^z}=1-\frac{Q^{2}}{2d^{2}r^{z}},\nonumber\\
A_{t}=\sqrt{\frac{2(z-1)}{z}}Lr^zf(r), ~~~  \mathcal{F}_{rt}=QLr^{z-d-1},~~~ m^{2}_{vector}=\frac{z d}{L}.
\end{eqnarray}
which is asymptotical to the Lifshitz spacetime as $r$ goes to infinity. Here $z$ and $Q$ are
dynamical exponent and charge parameter, respectively. $r_{+}$ denotes the position of horizon which
satisfies $f(r_{+})=0$. As discussed in \cite{DWP}, it is worth noting that the solution (\ref{blackhole})
is only valid for $z=2d>1$. We emphasize that the charged Lifshitz black hole is very different from the charged black holes in AdS spacetime because it does not admit extremal solutions. Thus
the behavior of charged Lifshitz black holes looks  more analogous
to Schwarzschild-AdS black holes than Reissner-Nordstr$\ddot{o}$m-AdS (RN-AdS) black holes.
From $\mathcal{F}_{\mu\nu}$ in the above solution (\ref{blackhole}), we can obtain the gauge field
\begin{equation}\label{At}
\mathcal{A}_{t}=\frac{QL}{d-z}(1-\frac{r_{+}^{d-z}}{r^{d-z}}).
\end{equation}
Note that when $z=1$ and $d=2$, (\ref{At}) reduces to the scalar potential of charged black holes. We stress that the vector field $\mathcal {A}_t$ is massless and should be used in the following discussions of holographic fermions. The metric is invariant under the Lifshitz scaling
\begin{equation}\label{}
t\rightarrow l^{-z}t,~~~l\rightarrow l r,~~~~ x_i\rightarrow l^{-1}x_i.
\end{equation}
The Hawking temperature at the event horizon is given by
\begin{equation}\label{temperature}
T=\frac{z}{4\pi}r_{+}^{z},
\end{equation}
where
\begin{equation}
r_{+}^{z}=\frac{Q^{2}}{2d^{2}}.
\end{equation}
The Hawking temperature is also invariant under the transformation $T\rightarrow l^z T$.
The entropy of the black hole is
\begin{equation}\label{entropy}
S_{BH}=\frac{L^d V_d}{4G_{d+2}}r_{+}^{d}
\end{equation}
with $V_d$ the volume of the d-dimensional spatial part.

In the rest of this work, we will apply the black hole solution described above to discuss some properties
of the holographic fermions in this background. Although the charged Lifshitz black hole has no extremal limit, we expect that at a low but finite temperature we can find evidence of Fermi surfaces.

\section{Dirac Equation}
In order to study fermions in the dual boundary theory, we consider the bulk action for a probe Dirac fermion with the mass $m$ and charge $q$.\footnote{The bulk action of fermion with bulk dipole coupling was discussed in \cite{dipole}.} The
bulk fermion action is
\begin{equation}
S_{bulk}=\int
d^{d+2}x\sqrt{-g}i\bar{\psi}\Big(\Gamma^a D_{a}-
m\Big)\psi,
\end{equation}
where $\Gamma^a=(e_\mu)^a\Gamma^\mu$ and the covariant derivative $D_{a}$ is
\begin{equation}
D_{a}=\partial_{a}+\frac{1}{4}(\omega_{\mu\nu})_a\Gamma^{\mu\nu}-iq\mathcal{A}_a,
\end{equation}
with $\Gamma^{\mu\nu}=\frac{1}{2}[\Gamma^\mu,\Gamma^\nu]$. Here, $a, b$ are usual spacetime abstract index and $\mu, \nu$ are the tangent-space index. We should notice that $\mathcal{A}_a$ correspond to the massless $U(1)$ gauge field  in (\ref{a}). The spin connection 1-forms
\begin{equation}\label{w}
(\omega_{\mu\nu})_{a}=(e_\mu)^b\nabla_a(e_\nu)_b.
\end{equation}
Here $(e_\mu)^a$ form a set of orthogonal normal vector bases.
The Dirac equation derived from the above action reads
\begin{equation}
\Gamma^a D_{a}\psi-m\psi=0.
\end{equation}
Without loss of generality, we choose the orthogonal normal vector bases as follows
\begin{eqnarray}
(e_t)^a=\sqrt{g^{tt}}(\frac{\partial}{\partial{t}})^a,\nonumber\\
(e_i)^a=\sqrt{g^{xx}}(\frac{\partial}{\partial{x^i}})^a,\nonumber\\
(e_r)^a=\sqrt{g^{rr}}(\frac{\partial}{\partial{r}})^a.
\end{eqnarray}
Then based on the (\ref{w}), the non-vanishing components of spin connections can be obtained as
\begin{eqnarray}
(\omega_{tr})_a=-(\omega_{rt})_a=-\sqrt{g^{rr}}\partial_r\sqrt{g_{tt}}(dt)_a,\nonumber\\
(\omega_{ir})_a=-(\omega_{ri})_a=-\sqrt{g^{rr}}\partial_r\sqrt{g_{xx}}(dx^i)_a.
\end{eqnarray}
In order to remove the spin connection in Dirac equation and to investigate in Fourier space,
following \cite{f2}, we make a transformation
\begin{equation}
\psi=(-g g^{rr})^{-\frac{1}{4}}e^{-i\omega t+ik_{i}x^{i}}\phi.
\end{equation}
Considering the rotation symmetry in the spatial directions, we can simply choose the momentum along the $x$-direction  $k=k_x$ . Then,
the Dirac equation can be written as
\begin{equation}\label{phieom}
\sqrt{g^{rr}}\Gamma^r\partial_r\phi-i(\omega+q\mathcal{A}_t)\sqrt{g^{tt}}\Gamma^t\phi+ik\sqrt{g^{xx}}\Gamma^x\phi-m\phi=0
\end{equation}
As in \cite{f3,IL,JPW2}, we will consider the following basis for gamma matrices
\begin{eqnarray}
\Gamma^r=\left(
\begin{array}{cc}
-\sigma^{3}\bold{1} & 0 \\
0 & -\sigma^3\bold{1} \\
\end{array}\right),
\ \ \Gamma^t=\left(
\begin{array}{cc}
 i\sigma^1\bold{1} & 0 \\
 0 & i\sigma^1\bold{1} \\
\end{array}\right),
\ \ \Gamma^x=\left(
 \begin{array}{cc}
 -\sigma^2\bold{1} & 0 \\
  0 & \sigma^2\bold{1}\\
\end{array} \right),
\cdots\ \
\end{eqnarray}
where $\bold{1}$ is an identity matrix which have size $2^{\frac{d-3}{2}}$ for $d$ odd and size $2^{\frac{d-4}{2}}$ for $d$ even. Now,we set $\phi=\left(
 \begin{array}{c}\phi_1 \\ \phi_2\\ \end{array} \right)$, then (\ref{phieom}) becomes
\begin{equation}\label{phi12eom}
\sqrt{g^{rr}}\partial_r\left(
 \begin{array}{c}\phi_1 \\ \phi_2\\ \end{array} \right)+m\sigma^3\otimes\left(
 \begin{array}{c}\phi_1 \\ \phi_2\\ \end{array} \right)=\sqrt{g^{tt}}(\omega+q \mathcal{A}_t)i\sigma^2\otimes\left(
 \begin{array}{c}\phi_1 \\ \phi_2\\ \end{array} \right)\mp k\sqrt{g^{xx}}\sigma^1\otimes\left(
 \begin{array}{c}\phi_1 \\ \phi_2\\ \end{array} \right)
\end{equation}
In order to decouple the equation of motion, we set $\phi_I=\left(\begin{array}{c}y_I \\ z_I\\\end{array} \right)$,
with $I=1,2$.  The equation of motion yields
\begin{equation}\label{eom}
(\sqrt{g^{rr}}\partial_{r}+2m)\xi_I=\Big[\sqrt{g^{tt}}(\omega+q\mathcal{A}_t)+(-1)^{I}k\sqrt{g^{xx}}\Big]
+\Big[\sqrt{g^{tt}}(\omega+q\mathcal{A}_t)-(-1)^{I}k\sqrt{g^{xx}}\Big]\xi_I^2,
\end{equation}
where $\xi_I=\frac{y_I}{z_I}$. Substituting the black hole sulution (\ref{blackhole}) to the Dirac equation and make some
algebraic calculation, we have
\begin{equation}\label{eomlifshitz}
(r\sqrt{f}\partial_{r}+2m)\xi_I=\Big[\frac{(\omega+q\mathcal{A}_t)}{r^z\sqrt{f}}+(-1)^{I}\frac{k}{r}\Big]
+\Big[\frac{(\omega+q\mathcal{A}_t)}{r^z\sqrt{f}}-(-1)^{I}\frac{k}{r}\Big]\xi_I^2.
\end{equation}
Asymptotically, when $r\rightarrow\infty$, namely towards the boundary, the solution of the Dirac equation (\ref{phi12eom}) can be written as
\begin{equation}\label{bdysol}
\phi_I\rightarrow A_I r^m\left(
\begin{array}{c}
0 \\1 \\ \end{array}
 \right)+B_Ir^{-m}\left(
\begin{array}{c} 1 \\ 0 \\\end{array}\right).
\end{equation}
If we suppose two coefficients in (\ref{bdysol}) are related by
\begin{equation}
B_I \left(
\begin{array}{c}
 1 \\0 \\\end{array}
 \right)= \mathcal{S}A_I
 \left(\begin{array}{c} 0 \\1 \\
\end{array}\right),
\end{equation}
then Green function on the boundary is\cite{IL}
\begin{equation}
G=-i\mathcal{S}\gamma^0.
\end{equation}
So the Green function can be written as
\begin{equation}\label{gi}
G=\lim_{r\rightarrow\infty}r^{2m}
    \left(
             \begin{array}{cc}
               \xi_1 & 0 \\
               0 & \xi_2 \\
             \end{array}
             \right).
\end{equation}
On the other hand, the ratio $G=\frac{B_{I}}{A_{I}}$ can be fixed by imposing the in-falling boundary condition for $\phi_{I}$ at the event horizon. where $\phi_{I}$ is behave as
\begin{equation}
\phi_{I}\propto \left(\begin{array}{c} i \\1 \\
\end{array}\right)e^{-i\omega r_{+}},
\end{equation}
with $r_{+}=\int\frac{dr}{r^2 f}$. That is to say, this ration can be written as
\begin{equation}
G=\lim_{r\rightarrow\infty}r^{2m} \xi_{I},
\end{equation} which is same as (\ref{gi}).
The in-falling boundary condition at the horizon yields
\begin{equation}\label{bdycond}
\xi_I\mid_{r=r_{+}}=i.
\end{equation}

\section{Numerical calculation}
In this section, we will explore some properties of this system by investigating the behavior of the retarded Green function. At first, we try to obtain the fermi
momentum  numerically in the Lifshitz background (\ref{blackhole}). Before processing, we would like to point out that by taking $k\rightarrow-k$,
the relation between $G_{11}$ and $G_{22}$ is $G_{22}(\omega,k)=G_{11}(\omega,-k)$ that can be deduced from
the equation of motion (\ref{eom}) and the boundary condition (\ref{bdycond}). Thus, we  only focus on $G_{22}$
in the following discussion. For convenience of our numerical calculation, we can set $m=0$, $L=1$.

\subsection{Charge dependence}

In condensed-matter physics one more often deals with the
situation that under a scale transformation the time direction scales twice as fast
as the spatial directions, i.e., the dynamical exponent is $z=2$. So we will discuss $z=2$ case with different charge $q$ in this subsection.
Here, we only study the system with finite temperature. At first, we will fix the temperature at $T=\frac{1}{4\pi}$. The behavior of $ImG_{22}$ can be summarized as follows:

\begin{enumerate}
\item
From Figure \ref{z2k} and Figure \ref{z23d}, we can see that in the limit $\omega\rightarrow0$, $ImG_{22}$  has a sharp quasiparticle peak at the Fermi momentum $k_F$, which agrees with \cite{f2,f4}. Note that in order to determine $k_F$ numerically, we have set $\omega=1\times10^{-12}$. Figure \ref{z2k} shows that $k_F\approx1.9$ for $q=2$, $k_F\approx2.5$ for $q=2.5$ and $k_F\approx3.2$ for $q=3$.\footnote{From the third part ($q=3$) in Figure 1, we can find there are two peaks in this case. As mentioned in \cite{f2}, when $q$ increased, new branches of Fermi surfaces appear (i.e. when $q=3$, the Fermi surface has two branches). Another peak is occurred at $k\approx-1.8$ . But, we only work on the peak located at $k\approx3.2$ in this subsection. } More accuracy, the sharp quasi-particle-like peak is located at $k_F=1.88553916$, $k_F=2.53153814$ and $k_F=3.18252078$ for $q=2$, $q=2.5$ and $q=3$, respectively. This implies the Fermi momentum will increase with the increasing of the charge $q$. This result is also consistent with the discussion in \cite{f2}. Our result also show that for smaller charge $q\leq 1$, there is no sharp peak at the temperature $T=\frac{1}{4\pi}$.

\begin{figure}
\centering
\includegraphics[width=.32\textwidth]{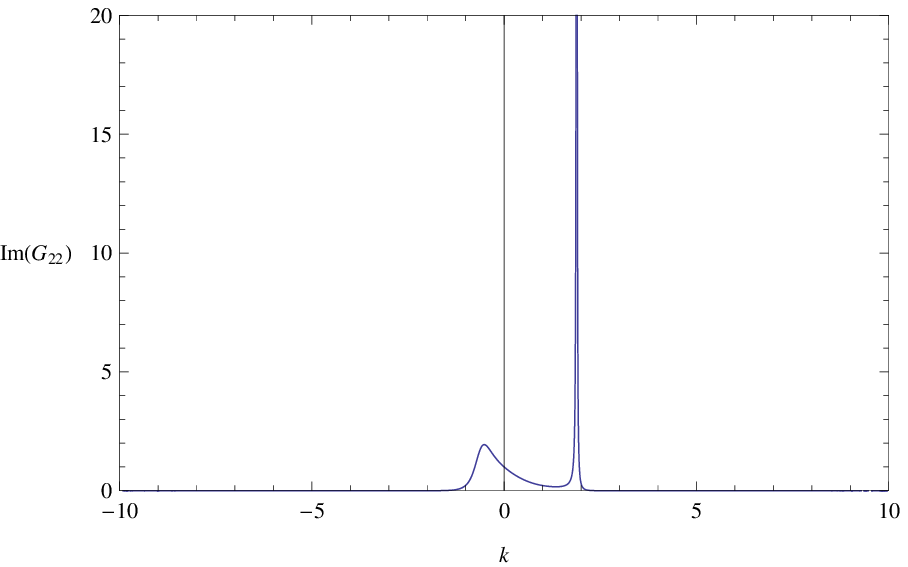}
\includegraphics[width=.32\textwidth]{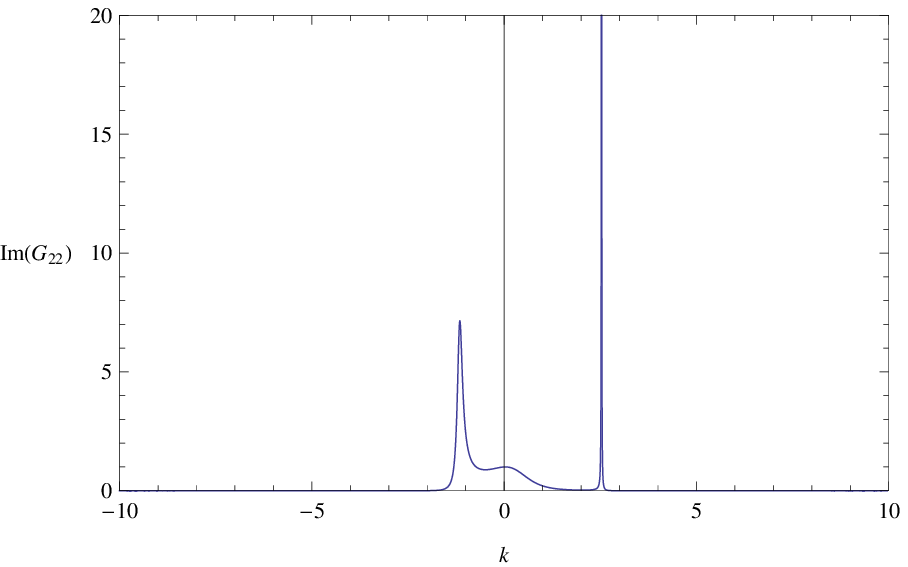}
\includegraphics[width=.32\textwidth]{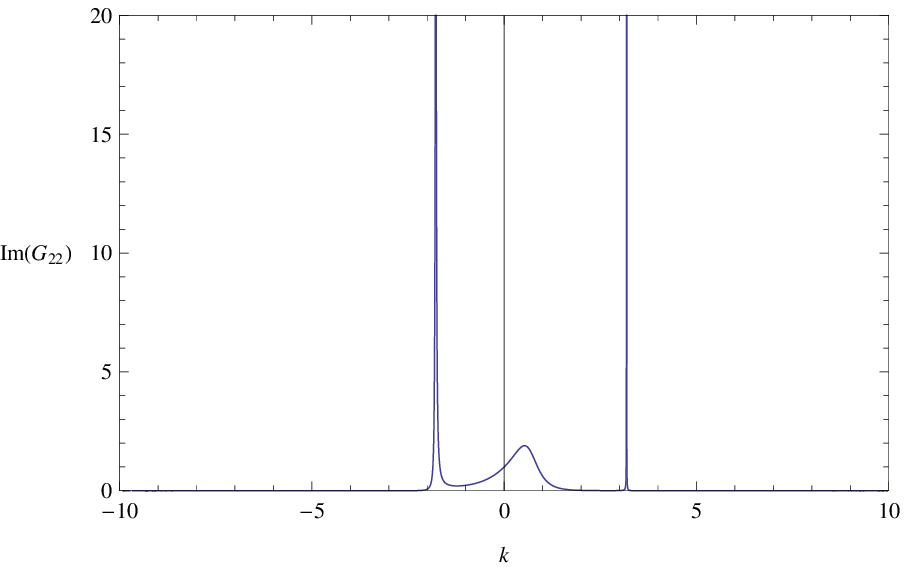}
\caption{The plot of $ImG_{22}(k)$ for $q=2$, $q=2.5$ and $q=3$ ($\omega=1\times10^{-12}$). }
\label{z2k}
\end{figure}

\item
 We present the 3D plot of $ImG_{22}(\omega,k)$ in Figure \ref{z23d}. It is obvious that the sharp quasi-particle-like peak occurs near $k_F$  for $q=2$, $q=2.5$ and $q=3$ from the three figures, respectively.
\begin{figure}
\centering
\includegraphics[width=.32\textwidth]{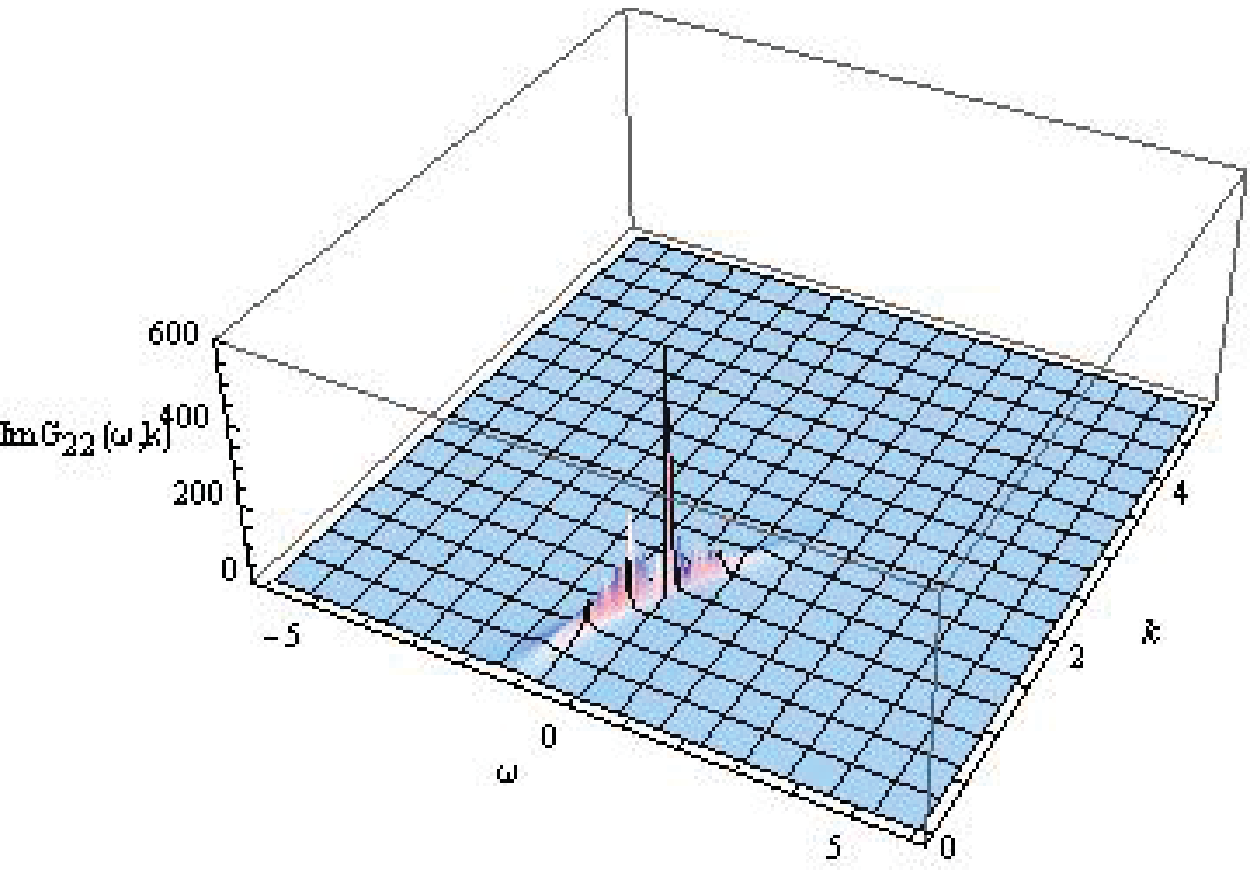}
\includegraphics[width=.32\textwidth]{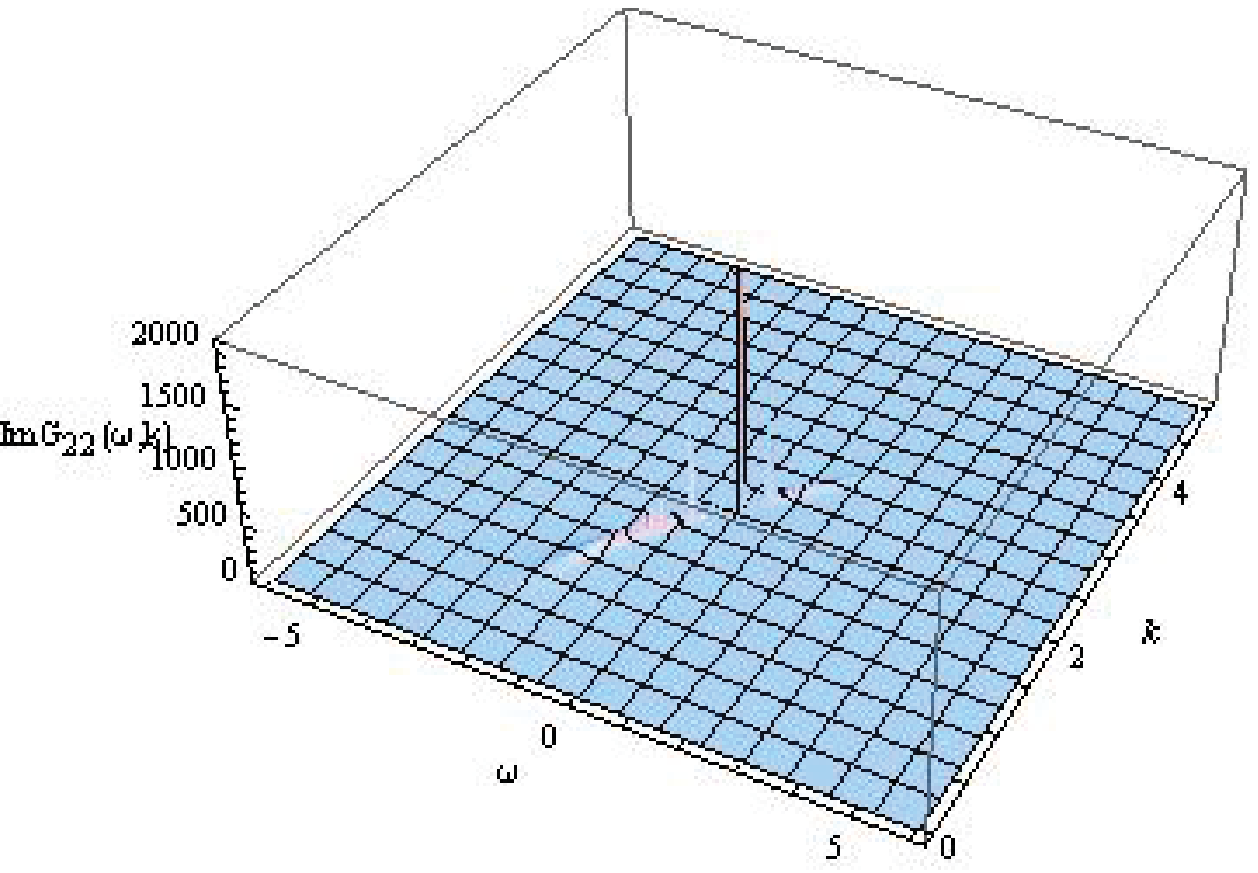}
\includegraphics[width=.32\textwidth]{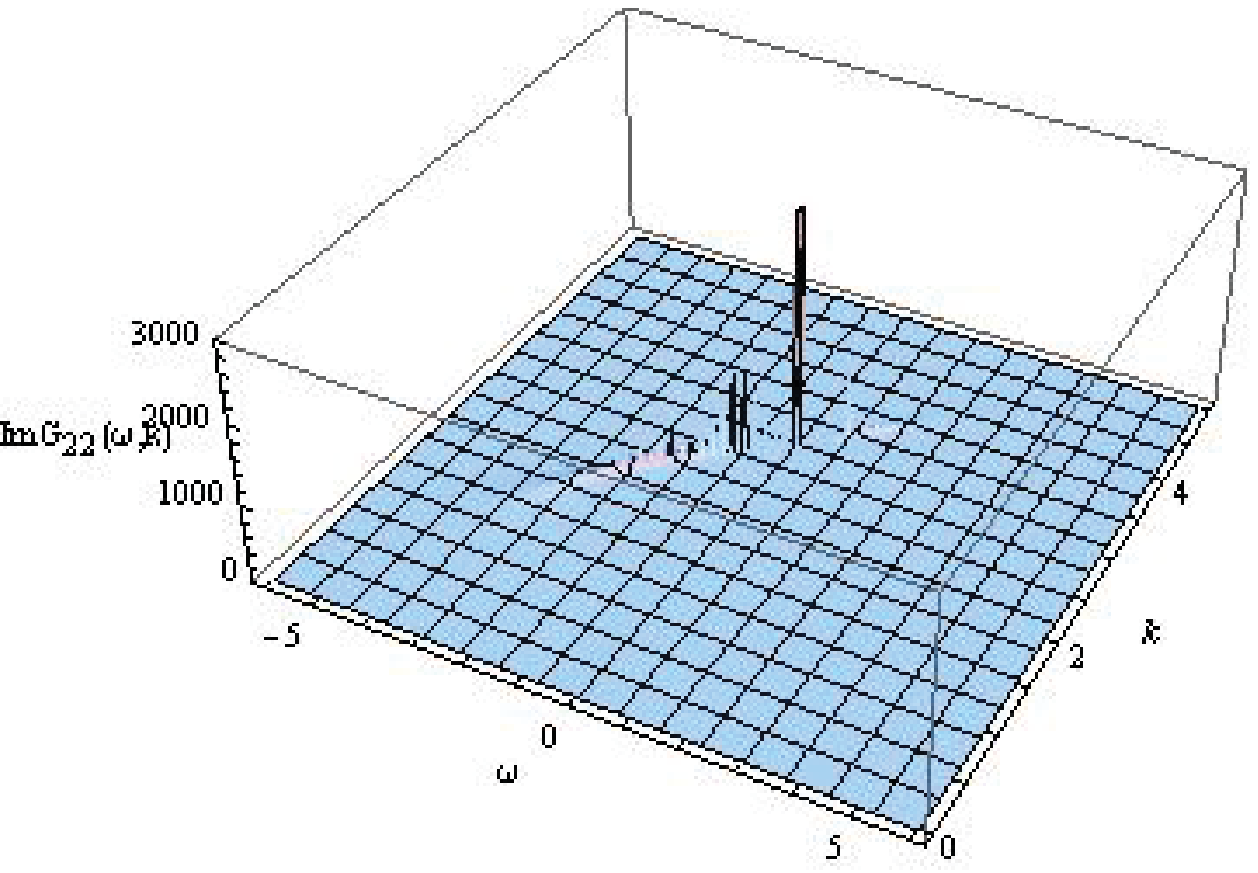}
\caption{The 3D plot for $q=2$, $q=2.5$ and $q=3$ ($\omega=1\times10^{-12}$). }
\label{z23d}
\end{figure}

\item
Based on the two points above, it is also of interest to explore the behavior of $ImG_{22}$ near the Fermi momentum
$k_{\perp}=k-k_F$. We show the dispersion relation
\begin{equation}\label{dr}
\omega_{\ast}(k_{\perp})\sim k_{\perp}^{\alpha}
\end{equation}
in Figure \ref{z2dr}. In the figure, the blue point is the data of $k_{\perp}$ and $\omega_{\ast}$. The red line is the plot
of fitting function of our data. Here we find the exponent $\alpha$ is always 1 by changing $q$ from 2, 2.5 to 3. In other words,
The linear dispersion relation will be hold for all $q$ in the charged Lifshitz black hole with $z=2$.\footnote{In \cite{f2}, the authors found that the dispersion relation exponent $\alpha$ decreases rapidly with increasing $q$. But here, we find that exponent $\alpha$ is  $1$ for different values of $q$. That is to say, the linear dispersion relation does not change in the $z=2$ charged Lifshitz background when $q$ increases. }
\begin{figure}
\centering
\includegraphics[width=.32\textwidth]{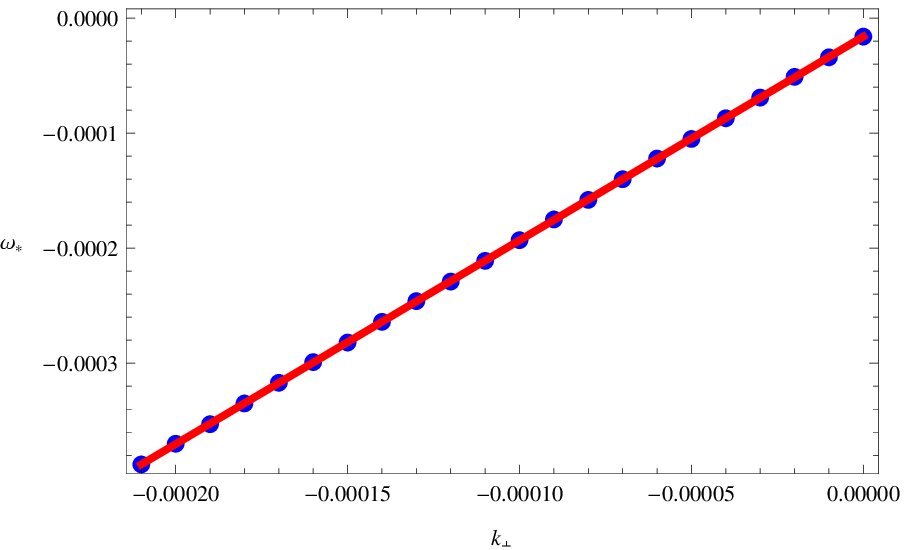}
\includegraphics[width=.32\textwidth]{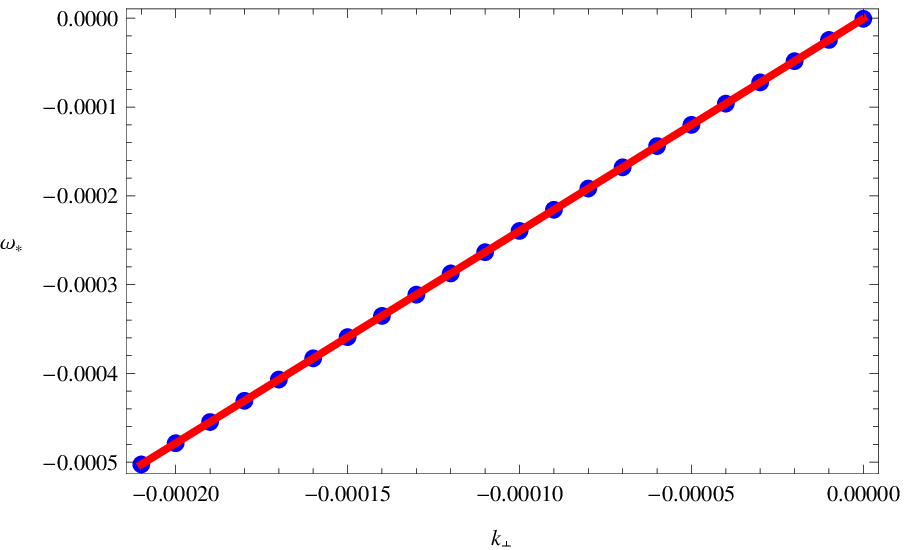}
\includegraphics[width=.32\textwidth]{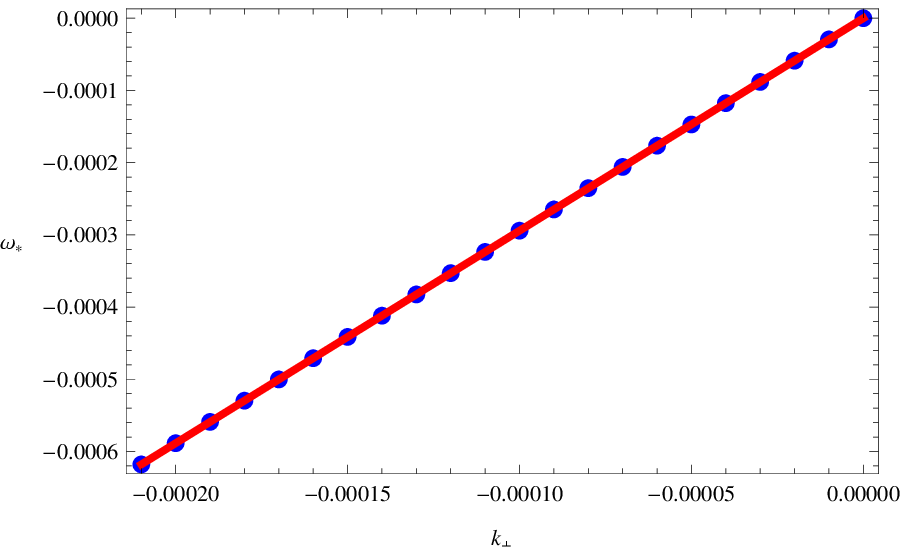}
\caption{The dispersion relation between $k_{\bot}$ and $\omega_{\ast}$ for $q=2$, $q=2.5$ and $q=3$. }
\label{z2dr}
\end{figure}

\item
However, to further investigate the dual liquid, we move on to check the relation between the maximum height of $ImG_{22}$ and $k_{\perp}$ (see Figure \ref{z2sccaling})
\begin{equation}\label{Gk}
ImG_{22}(\omega_{\ast}(k_{\perp}),k_{\perp})\sim k_{\perp}^{-\beta}
\end{equation}
In Figure \ref{z2sccaling}, the blue point is the data of $ImG_{22}(\omega_{\ast}(k_{\perp}),k_{\perp})$, and the red line is
the plot of the fitting function of the data. Interestingly, though $k_F$  drastically change with $q$,  we find here $\beta$ is around $2$ that is almost independent of the charge $q$. This phenomenon is also observed in \cite{f2}.
\begin{figure}
\centering
\includegraphics[width=.32\textwidth]{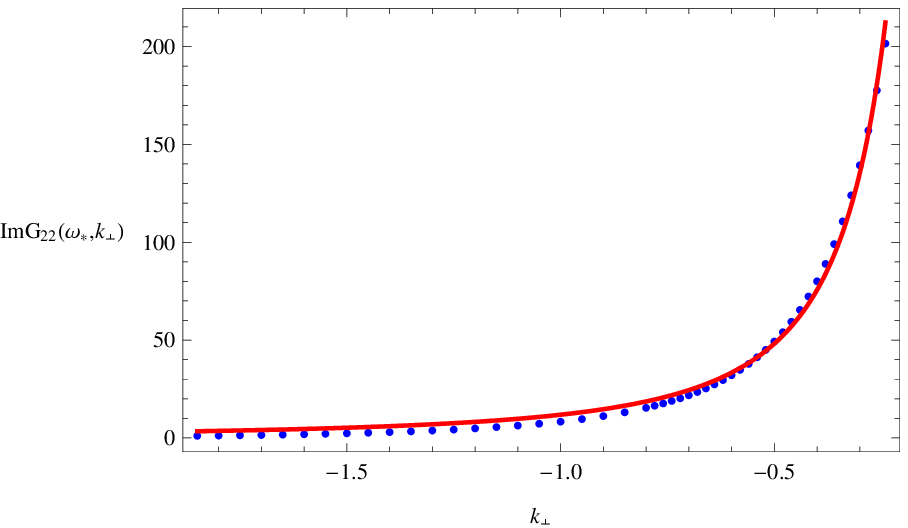}
\includegraphics[width=.32\textwidth]{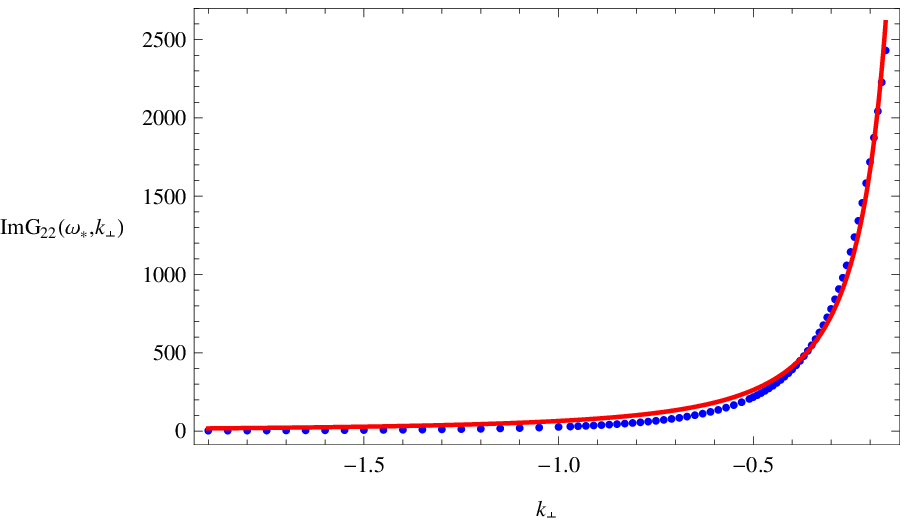}
\includegraphics[width=.32\textwidth]{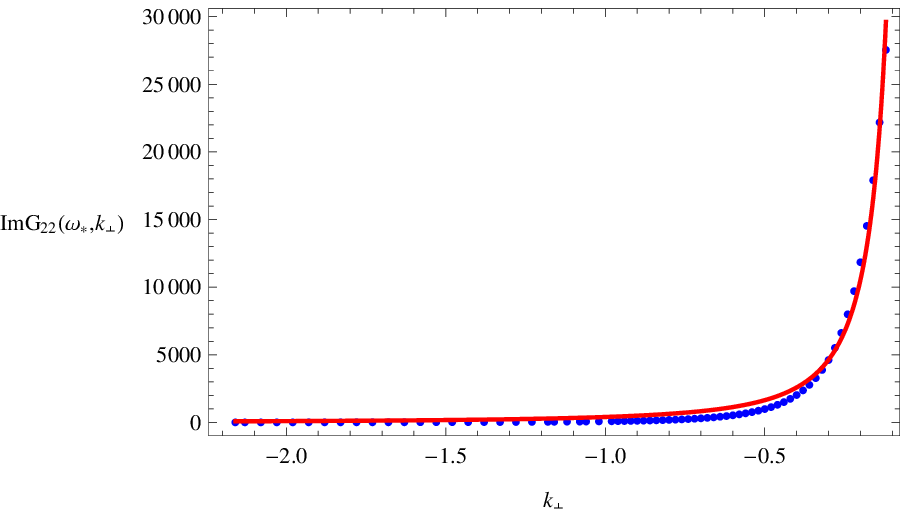}
\caption{The scaling behavior of the height of $ImG_{22}(\omega_{\ast}(k_{\perp}),k_{\perp})$ at the maximum for $q=2$, $q=2.5$ and $q=3$. }
\label{z2sccaling}
\end{figure}

\item
In Figure \ref{z2fit}, we examine the correctness of $\alpha$ and $\beta$ obtained above by checking the scaling behavior near the Fermi surface through fitting the whole curves of $G_{22}(\omega)$. For $q=2$, $q=2.5$ and $q=3$, we set $k=1.85$, $k=2.52$ and $k=3.18$, respectively. The fitting function of $G_{22}$ for both sign of $\omega$ is
\begin{equation}\label{ffun}
G_{22}(\omega,k_{\perp})\approx \frac{a_0 (-k_{\perp})^{-\beta}}{(\frac{-\omega}{a_1 (-k_{\perp})^{2\alpha}})-e^{i \gamma}}
\end{equation}
where $a_0$, $a_1$ and $\gamma$ are positive constants. The above function has a pole in the lower half $\omega$-plane at
\begin{equation}\label{wc}
\omega_c = -a_1 (-k_{\perp})^{2\alpha} e^{i \gamma}
\end{equation}
For (\ref{ffun}), $Re~\omega_c$ gives the location $\omega_{\ast}(k_{\perp})$ of the peak and $-Im~\omega_c$ determines the width $\Gamma$ of the peak,
\begin{equation}\label{decayrate}
\Gamma= (\tan{\gamma})~\omega_{\ast}^{2}(k_{\perp})
\end{equation}
The relation between $\Gamma$ and $\omega_{\ast}$  may  be the  reminiscent of the behavior of Fermi Liquid. We emphasize that the function (\ref{ffun}) is the best numerical fits we could find and from which we could qualitatively check that the scaling exponent $\alpha$ and $\beta$ obtained in Figure \ref{z2sccaling} is correct. \begin{figure}
\centering
\includegraphics[width=.32\textwidth]{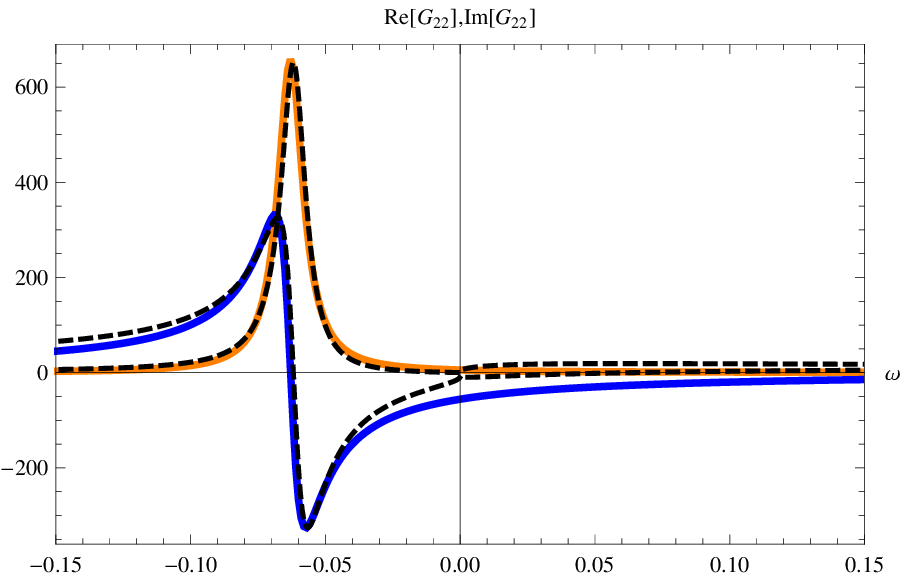}
\includegraphics[width=.32\textwidth]{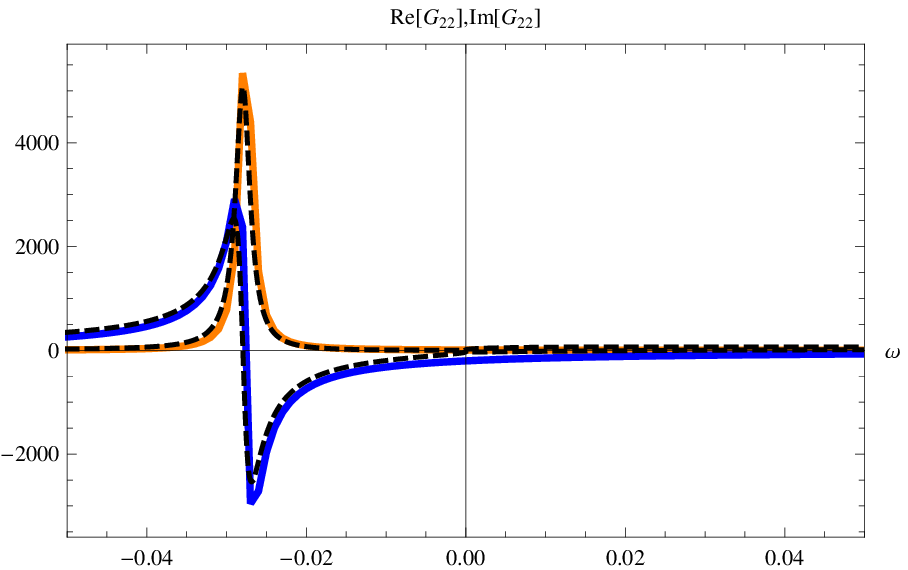}
\includegraphics[width=.32\textwidth]{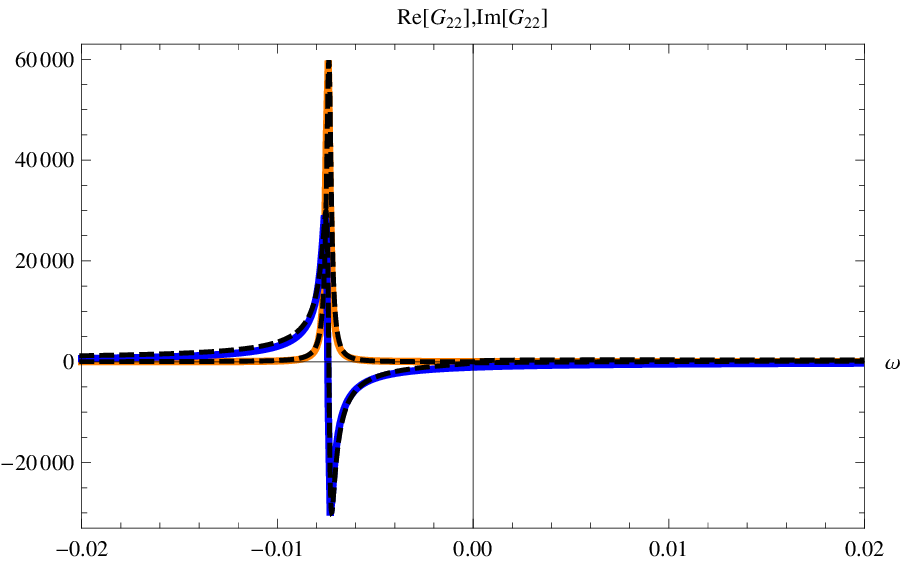}
\caption{The plot of $ReG_{22}(\omega)$ (blue), $ImG_{22}(\omega)$ (orange) and fit function (dashed lines) for $q=2$ with $k=1.85$ (left), $q=2.5$ with $k = 2.52$ (middle) and $q=3$ with $k = 3.18$ (right).}
\label{z2fit}
\end{figure}

\item
Up to now, we have examined the behavior of $Im G_{22} $ at a low temperature. In Figure \ref{T}, we compare the spectral function at different temperatures. As expected, the high temperature $T$ appears to smooth the peak due to thermal fluctuations, which agrees with \cite{f1}. If the temperature is high enough, the  sharp zero energy peak completely disappears.
\begin{figure}
\centering
\includegraphics[width=.32\textwidth]{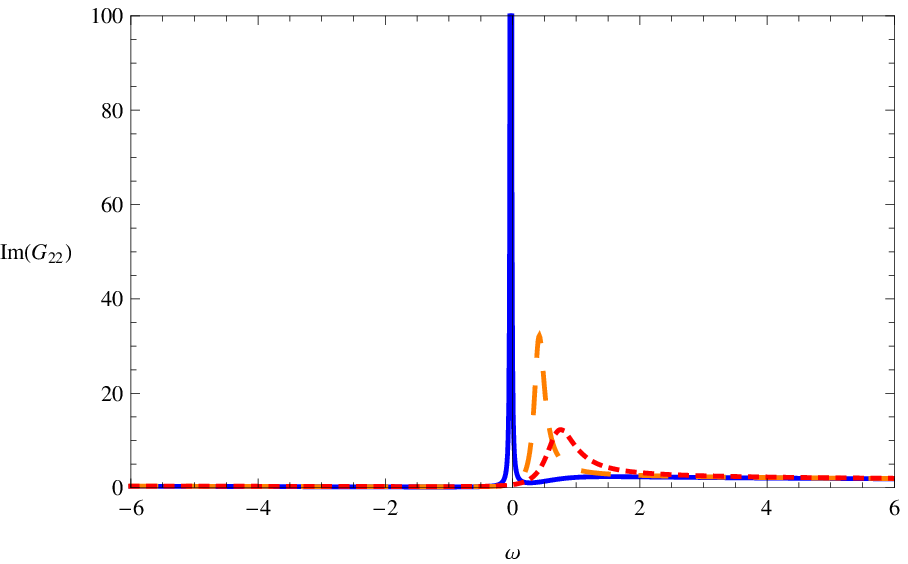}
\includegraphics[width=.32\textwidth]{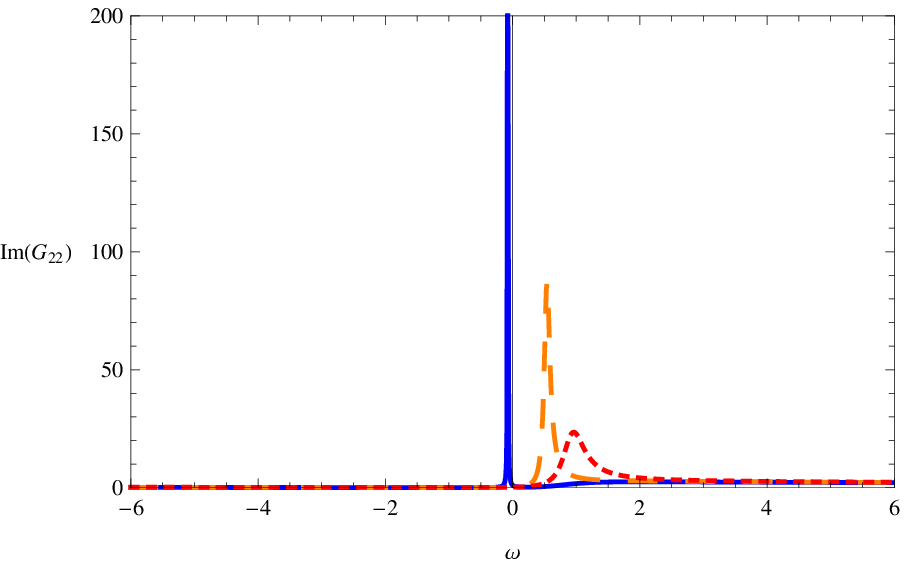}
\includegraphics[width=.32\textwidth]{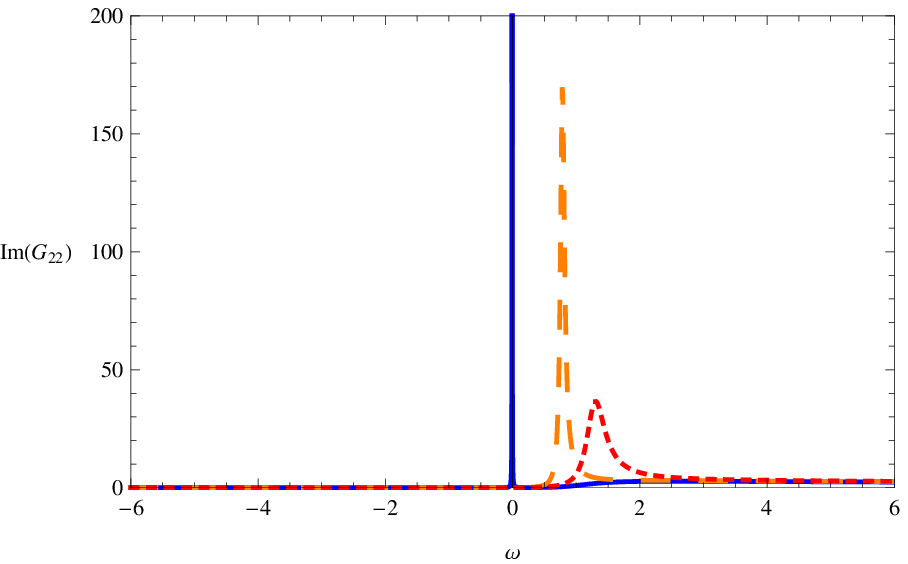}
\caption{The plot of $ImG_{22}(\omega)$ for $q=2(k=1.87)$, $q=2.5(k=2.5)$ and $q=3(k=3.18)$ with $T = \frac{1}{4\pi}$ (solid line), $T = \frac{2}{4\pi}$ (dashed line) and $T = \frac{3}{4\pi}$ (dotted line). }
\label{T}
\end{figure}
\end{enumerate}

\begin{table}
\centering
\begin{tabular}{|c|c|c|c|c|c|}
  \hline
  $q$ & 2 & 2.5 & 3 & 3.5 & 4 \\ \hline
  $k_F$ & 1.88553916 & 2.53153814 & 3.18252078 & 3.83819011 & 4.49787005 \\ \hline
  $\alpha$ & 1 & 1 & 1 & 1 & 1 \\ \hline
  $\beta$ & 2.02086 & 2.01415 & 2.02449 & 2.01559 & 2.02255 \\ \hline
\end{tabular}
\caption{The Fermi momentum and scaling behaviors with different $q$}
\label{fms}
\end{table}
In Table \ref{fms},  we summarize the  results for the Fermi momentum and scaling behaviors. We  observe that the Fermi momentum $k_F$ increases almost linearly as $q$ increases, while the scaling parameters $\alpha$ and $\beta$ approximately independent of  $q$.
According to \cite{Senthil1,Senthil2}, Landau Fermi liquid is characterized by the scaling behaviors $\alpha=\beta=1$. However, in our case the exponent $\beta\simeq2$ although $\alpha=1$.
From the expression (\ref{decayrate}), it seems that this system characterized by the linear dispersion and the quasi-particle width quadratic in frequency resembles the Landau Fermi liquid. This is simply because  the scaling form is destroyed and  Senthil's theory does not apply to our case \cite{Senthil1}. Let us explain this point more clearly.
It is well known that  the standard Fermi Liquid form is
\begin{equation}\label{flg}
G(k,\omega)= \frac{1}{\omega-v_F (k-k_F) +i \Gamma}
\end{equation}
where the width $\Gamma\sim\omega^2$. At first glance, the spectral function yields $A(k,\omega)\sim\frac{1}{(k-k_F)^2}$, which may give us $\beta=2$. However, the proper scaling form for Fermi Liquid is a delta function: $A(k,\omega)\sim \delta(k-k_F-\omega)$ and $\beta=1$ for this function. A more important parameter to characterize the property of the system is the peak width $\Gamma$ \cite{f2} as given in \ref{decayrate}. Thus, our holographic system preserves two features, i.e., the linear dispersion and quadratic quasi-particle width, of Fermi liquid at finite temperature. However, another important constraint for Landau Fermi liquid is Luttinger's theorem which states that the charge density is equal to the volume enclosed by the Fermi surface.
After repeating the similar calculation in \cite{LHLuttinger} in which a modified Luttinger theorem is given, we find that our system explicitly violates the standard Luttinger's theorem. This is because that in the probe limit, the charge density is much larger than the volume enclosed by the Fermi surfaces.  So our holographic system can not be well described by the standard Fermi liquid theory. The similar holographic system of non-Fermi liquid with the linear dispersion and  the growth of the peak with $k_{\perp}$ is nonlinear also can be seen in \cite{JPW2}. The Luttinger's theorem can be satisfied when the back-reaction of the fermions is taken into account \cite{lif90,sch,mc}.

\subsection{Dynamical exponent $z$ dependence}
\begin{figure}
\centering
\includegraphics[width=.66\textwidth]{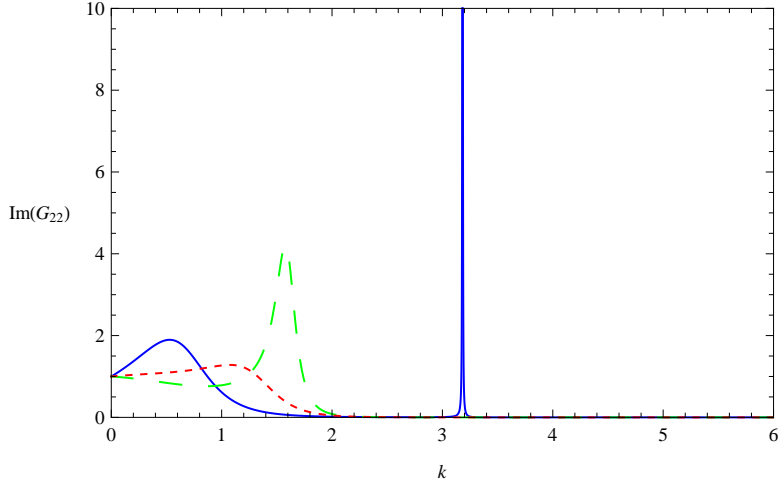}
 \caption{The plot of $ImG_{22}(k)$ for at $T=\frac{1}{4\pi}$ for $z=2$(solid line), $z=4$(dashed line) and $z=6$(dotted line). The peaks become smooth as $z$ increases ($\omega=1\times10^{-12}$). }
 \label{q3zk}
\end{figure}

\begin{figure}
\centering
\includegraphics[width=.48\textwidth]{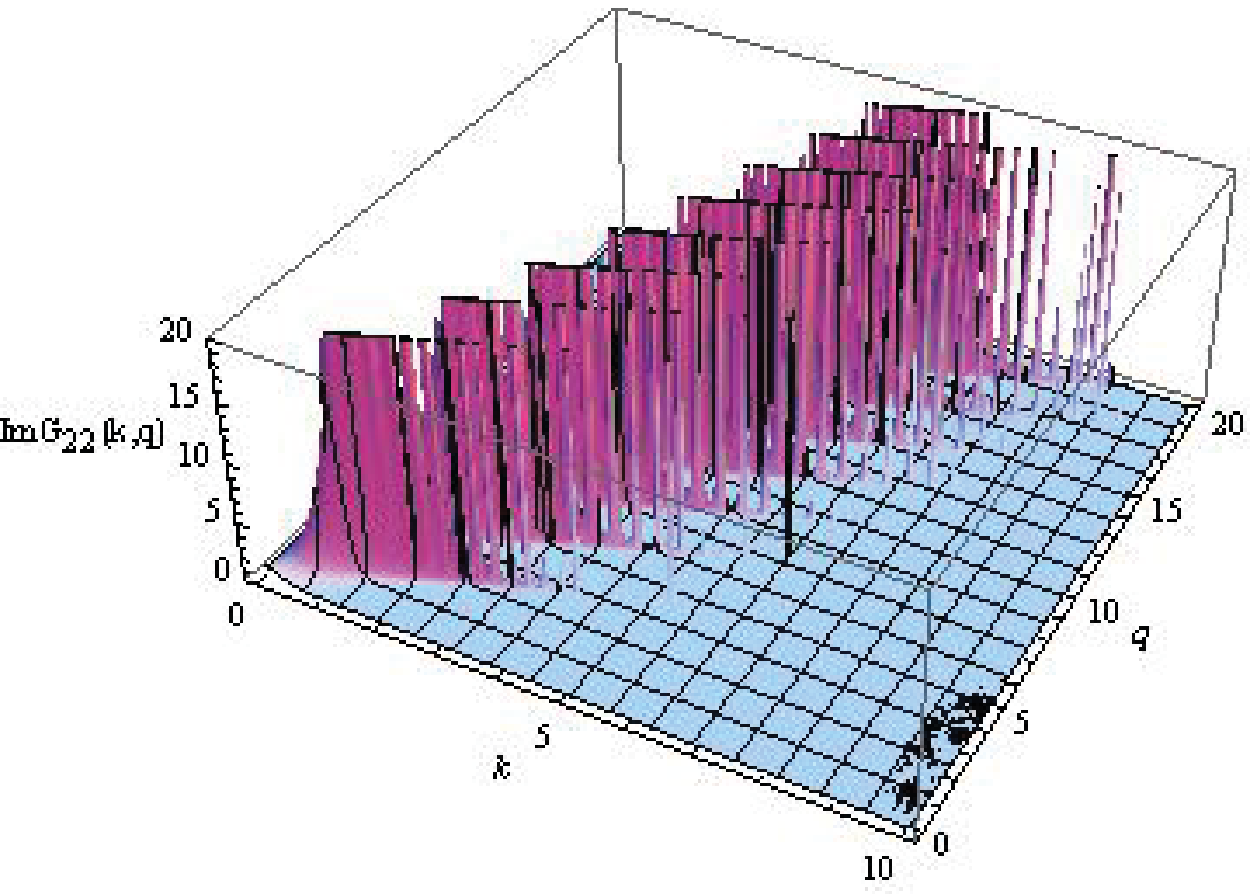}
\includegraphics[width=.39\textwidth]{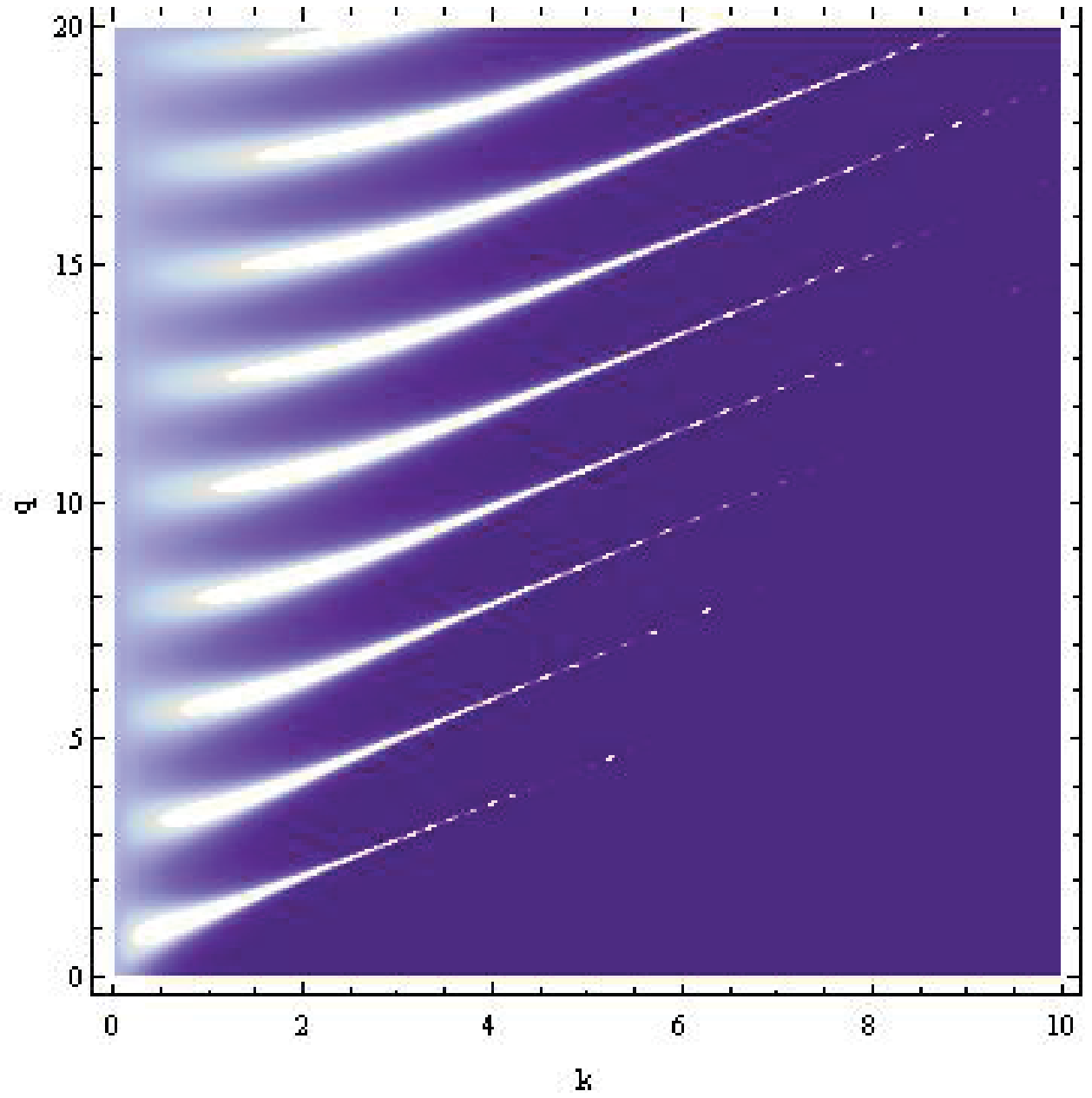}
\caption{The 3D plot of $ImG_{22}(k,q)$ and its density plot for $z=2$ at $T=\frac{1}{4\pi}$ ($\omega=1\times10^{-12}$). }
\label{z2kq}
\end{figure}

\begin{figure}
\centering
\includegraphics[width=.48\textwidth]{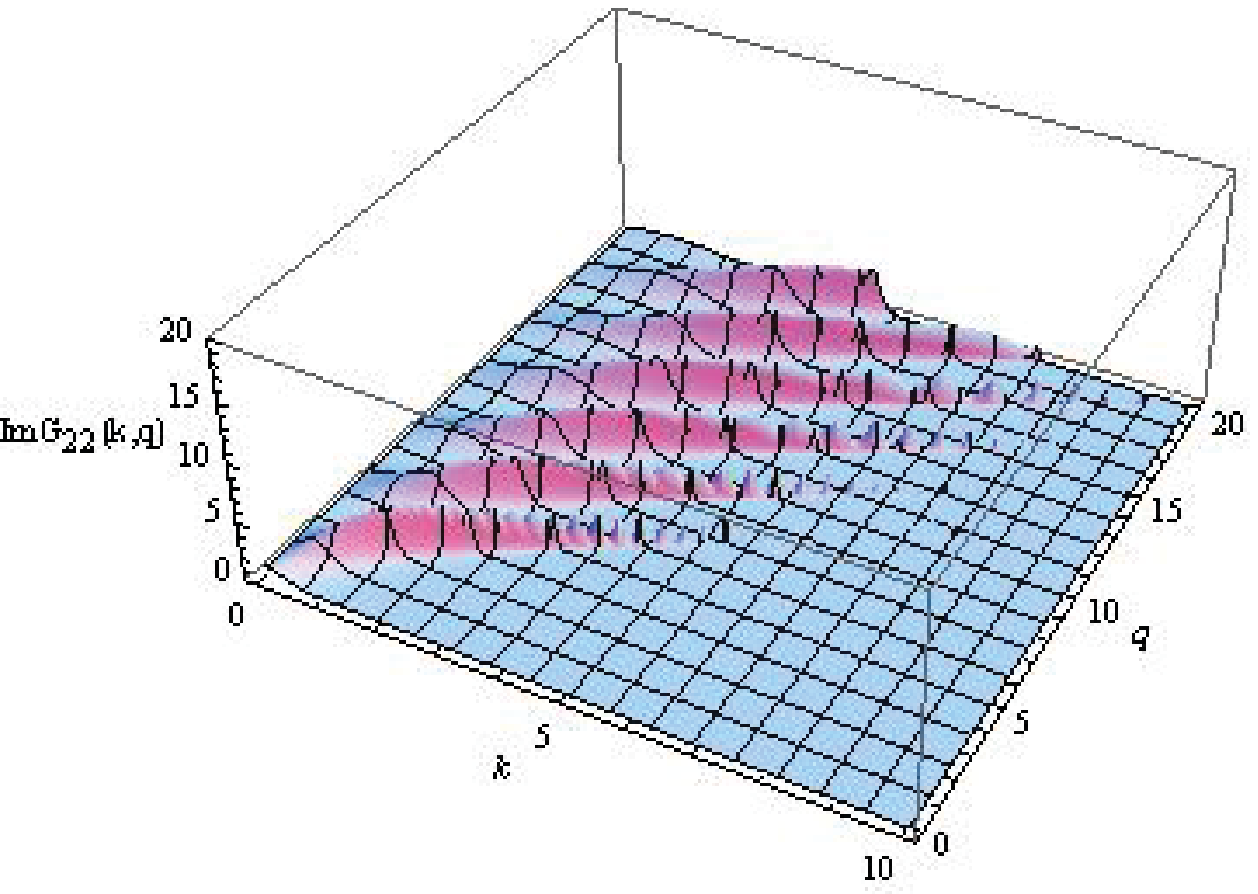}
\includegraphics[width=.39\textwidth]{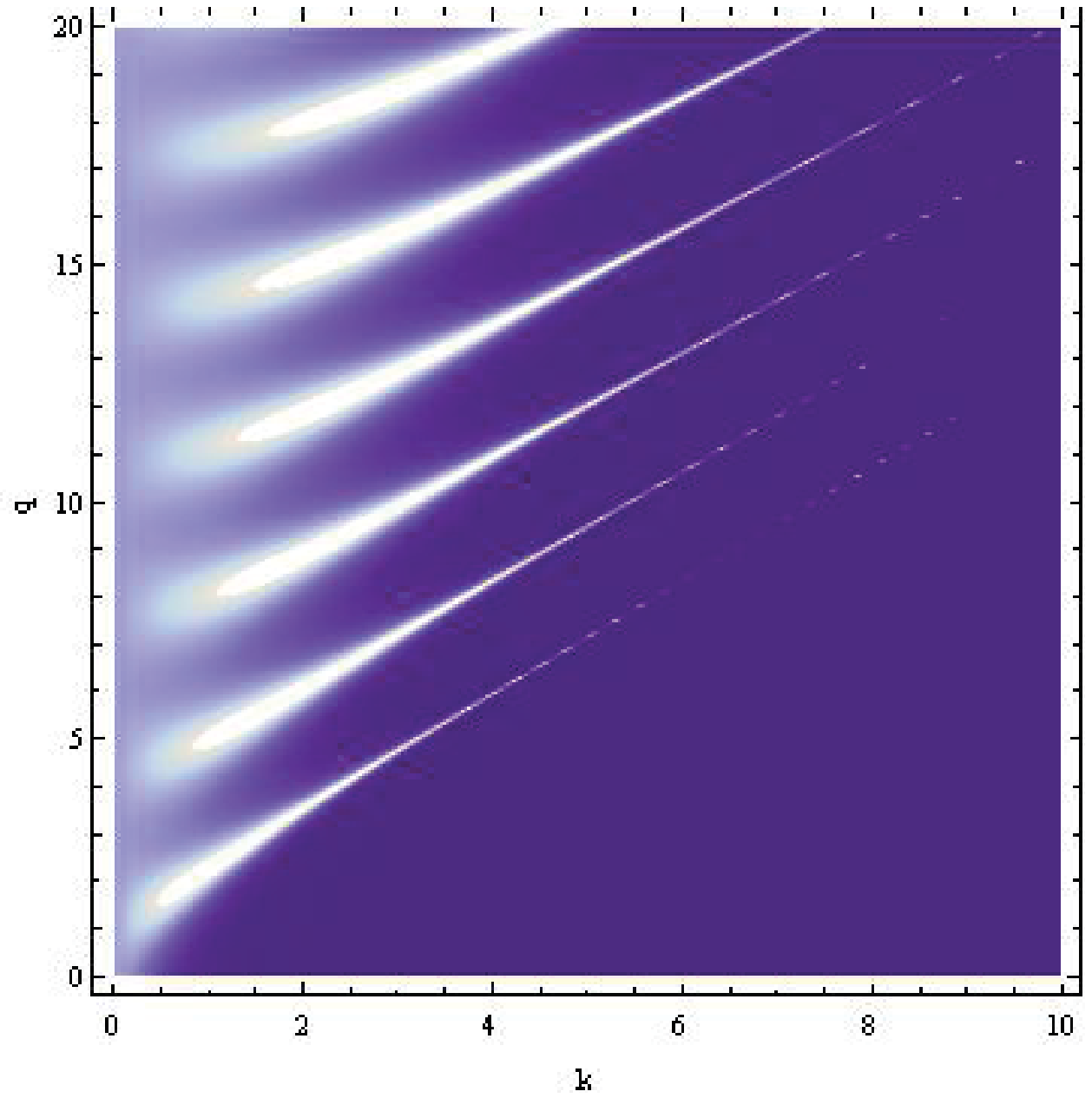}
\caption{The 3D plot of $ImG_{22}(k,q)$ and its density plot for $z=4$ at $T=\frac{1}{4\pi}$ ($\omega=1\times10^{-12}$). }
\label{z4kq}
\end{figure}

\begin{figure}
\centering
\includegraphics[width=.48\textwidth]{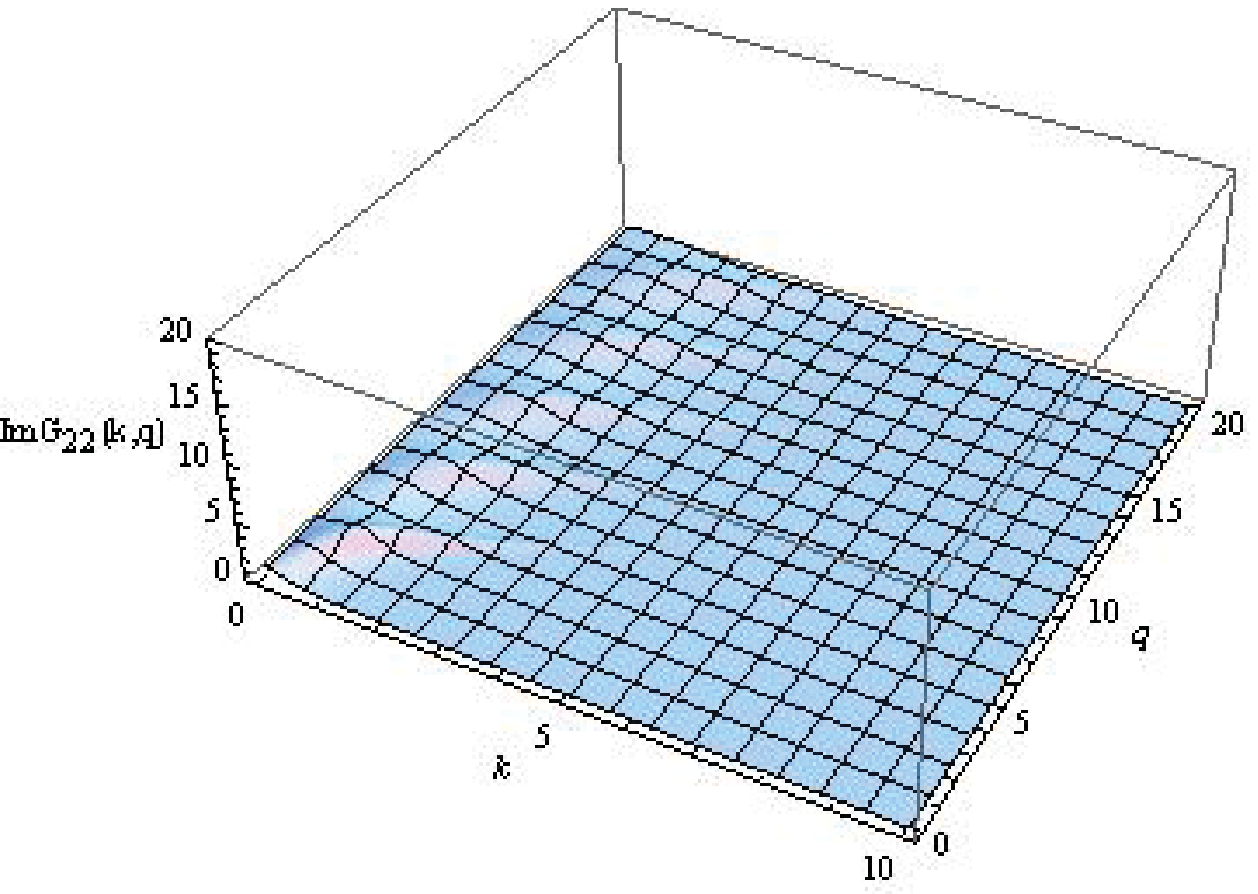}
\includegraphics[width=.39\textwidth]{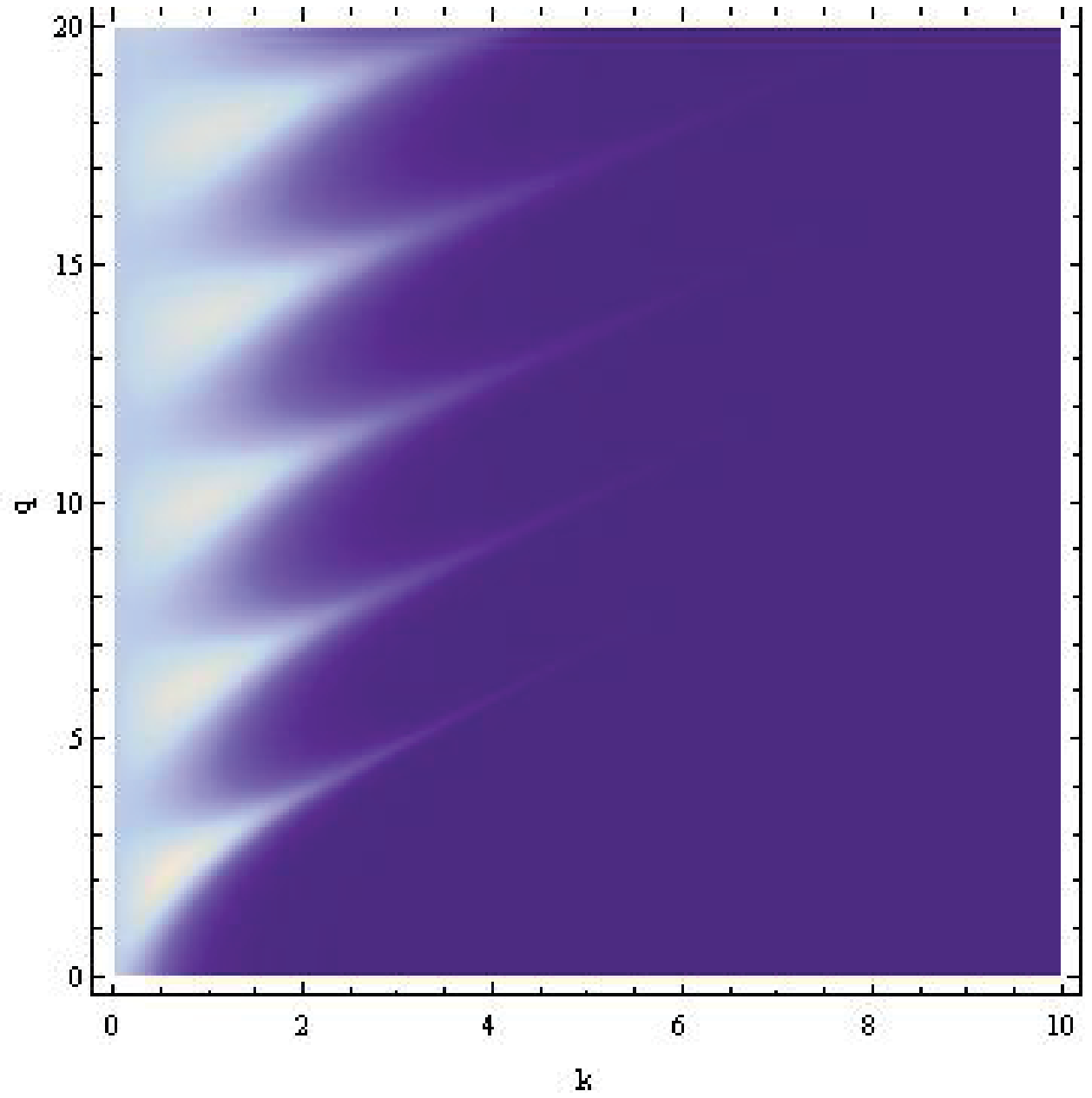}
\caption{The 3D plot of $ImG_{22}(k,q)$ and its density plot for $z=6$ at $T=\frac{1}{4\pi}$ ($\omega=1\times10^{-12}$). }
\label{z6kq}
\end{figure}

In this subsection, we study the influence of dynamical exponent $z$ on the behavior of the retarded  Green function. We know that for fixed $r_{+}$, the Hawking temperature $T=\frac{z}{4\pi}r^z_{+}$ increases as $z$ increases.  As shown in Figure \ref{T}, the high temperature appears to smooth the peak due to thermal fluctuations.

To have a clear picture on the effect of the dynamical exponent $z$, we would like to fix the temperature $T$ instead and let $r_{+}$ vary as $z$ increases.
For comparison, we will study three cases i.e., $z=2,z=4$ and $z=6$. In our numerical calculation, we  fix $q=3$, $\omega=1\times10^{-12}$ and the temperature $T=\frac{1}{4\pi}$. Figure \ref{q3zk} shows that the height of $ImG_{22}$ becomes lower as $z$ is increased. For $z=6$, the sharp quasi-particle-like peak nearly disappear.

 Figure \ref{z2kq} shows us that when $z=2$ the quasi-particle peaks exist at different value of $q$. Actually, the Fermi momentum $k_{F}$ increases as q increases.
 The effects of $z$ can be seen clearly from Figures \ref{z4kq} and \ref{z6kq}: For $z=4$ and $z=6$ cases, the peak and the Fermi momenta are seriously suppressed no matter how big $q$ is. That is to say, in higher dimensional
 spacetime and for larger dynamical exponent $z$, the exponent $z$ plays a dominant role. As $z$ is large enough, the peak will be smoothed out.
 Meanwhile, we also find from the density plot of the three figures  that the spacing between the peaks  increases with $z$.

\section{Conclusion}
In this paper, we studied the properties of holographic fermions in charged Lifshitz black holes at low temperature for
massless fermions. In $z=2$ charged Lifshitz background, we first solved the Dirac equation numerically for three different cases(i.e. $q=2$, $q=2.5$ and $q=3$). For $q=2$, the sharp quasi-particle-like peak is located at
$k_F=1.88553916$.  For $q=2.5$, Fermi momentum of the sharp Fermi surface $k_F=2.53153814$. For $q=3$, Fermi momentum $k_F=3.18252078$. The value of the $k_F$ actually is very sensitive to the boundary conditions. When we increase $q$ ,the Fermi momentum $k_F$ increases with $q$ almost linearly.
As $q$ increased, the scaling parameter $\alpha$ remains to be $1$. We clarified that the scaling exponent $\beta$ ($\beta\approx2$) obtained here does not satisfy the scaling theory by Senthil \cite{Senthil1} and the interpretation of $\beta$ should be careful. Further  numerical computation leads to the conclusion that this system is characterized by the linear dispersion and quadratic quasi-particle width which are two important features of Landau Fermi liquid theory. However, the system is in fact violating Luttinger's theorem. This implies that the holographic system in this charged Lifshitz theory is actually fractionalized non-Fermi liquid\cite{LHLuttinger}.
 We then noticed that as the temperature increases, the singular zero
energy peak is smoothed down due to thermal fluctuations. After this, we find the peak of $ImG$ will smooth out when $z$ increases, and the physics behind this phenomena need us to further study. It would be interesting to include the magnetic field and compare the results of holography with real experiments data in the charged Lifshitz backgrounds in the future.

\section*{Acknowledgements}
The authors would like to thank  Yi Ling,  Da-Wei Pang,  Wei-Jia Li,  Yu Tian, Jian-Pin Wu, Xiao-Ning Wu and Hongbao Zhang  for useful discussions.  The work  was partly supported by NSFC (No. 11005072).  XHG  was also partly supported by Shanghai Rising-Star Program and Shanghai
Leading Academic Discipline Project (S30105).


\begin{thebibliography}{10}
\bibitem{adscft}
J. M. Maldacena,`` The Large N Limit of Superconformal Field Theories and Supergravity," Int. J. Theor. Phys. 38, 1113 (1999),
{[arXiv:hep-th/9711200]}.

\bibitem{gkp}
S. S. Gubser, I.R. Klebanov and A.M. Polyakov, ``Gauge Theory Correlators from Non-Critical String Theory," Phys.\ Lett.\
B 428 (1998) 105, {[arXiv:hep-th/9802109]}.

\bibitem{w}
E. Witten, ``Anti De Sitter Space And Holography," Adv.\ Theor.\ Math.\ Phys.\ {\bf 2} (1998) 253,
{[arXiv:hep-th/9802150]}.

\bibitem{f1}S. S. Lee, ``A Non-Fermi Liquid from a Charged Black Hole; A Critical Fermi Ball,"
Phys. Rev. D 79 (2009) 086006,
[arXiv:0809.3402 [hep-th]].
\bibitem{f2} H. Liu, J. McGreevy and D. Vegh, ``Non-Fermi liquids from holography,"
 Phys. Rev. D 83 (2011) 065029,
[arXiv:0903.2477 [hep-th]].
\bibitem{f3} T. Faulkner, H. Liu, J. McGreevy and D. Vegh, ``Emergent quantum criticality, Fermi surfaces, and AdS2," Phys. Rev. D 83 (2011) 125002,
[arXiv:0907.2694 [hep-th]].
\bibitem{f4} M. Cubrovic, J. Zaanen and K. Schalm, ``String Theory, Quantum Phase Transitions and the Emergent Fermi-Liquid," Science 325 (2009) 439,
[arXiv:0904.1993 [hep-th]].
\bibitem{lif0}S. Kachru, X. Liu and M. Mulligan, ``Gravity Duals of Lifshitz-like Fixed Points,"
Phys. Rev. D 78, 106005 (2008),
[arXiv:0808.1725 [hep-th]].
\bibitem{lif1} U. H. Danielsson and L. Thorlacius, ``Black holes in asymptotically Lifshitz spacetime," JHEP 0903, 070 (2009),
[arXiv:0812.5088 [hep-th]].
\bibitem{lif2} R. B. Mann, ``Lifshitz Topological Black Holes," JHEP 0906, 075 (2009),
[arXiv:0905.1136 [hep-th]].
\bibitem{lif3} G. Bertoldi, B. A. Burrington and A. Peet, ``Black Holes in asymptotically Lifshitz
spacetimes with arbitrary critical exponent,"
[arXiv:0905.3183 [hep-th]].
\bibitem{lif4} G. Bertoldi, B. A. Burrington and A. W. Peet, ``Thermodynamics of black branes in
asymptotically Lifshitz spacetimes," Phys. Rev. D 80, 126004 (2009),
[arXiv:0907.4755 [hep-th]].
\bibitem{lif5} M. Taylor, ``Non-relativistic holography,"
[arXiv:0812.0530 [hep-th]].
\bibitem{lif6} D. W. Pang, ``A Note on Black Holes in Asymptotically Lifshitz Spacetime,"
[arXiv:0905.2678 [hep-th]].\\
K. Balasubramanian and J. McGreevy, ``An analytic Lifshitz black hole,"
[arXiv:0909.0263 [hep-th]].\\
E. Ayon-Beato, A. Garbarz, G. Giribet and M. Hassaine, ``Lifshitz Black Hole in
Three Dimensions," Phys .Rev .D 80, 104029 (2009),
[arXiv:0909.1347 [hep-th]].\\
R. G. Cai, Y. Liu and Y. W. Sun, ``A Lifshitz Black Hole in Four Dimensional R2
Gravity," JHEP 0910, 080 (2009),
[arXiv:0909.2807 [hep-th]].\\
S. J. Sin, S. S. Xu and Y. Zhou, ``Holographic Superconductor for a Lifshitz fixed
point," International Journal of Modern Physics A, Volume: 26, Issue: 26(2011) pp. 4617-4631,
[arXiv:0909.4857 [hep-th]].\\
Y. S. Myung, Y. W. Kim and Y. J. Park, ``Dilaton gravity approach to three dimensional Lifshitz black hole," Eur. Phys. J. C70: 335-340, 2010
[arXiv:0910.4428 [hep-th]].\\
K. B. Fadafan, ``Drag force in asymptotically Lifshitz spacetimes",
[arXiv:0912.4873 [hep-th]].\\
J. Matulich, R. Troncoso, ``Asymptotically Lifshitz wormholes and black holes for Lovelock gravity in vacuum,"
[arxiv:1107.5568 [hep-th]].\\
\bibitem{lif7} T. Azeyanagi, W. Li and T. Takayanagi, ``On String Theory Duals of Lifshitz-like
Fixed Points," JHEP 0906, 084 (2009),
[arXiv:0905.0688 [hep-th]].
\bibitem{lif8} W. Li, T. Nishioka and T. Takayanagi, ``Some No-go Theorems for String Duals
of Non-relativistic Lifshitz-like Theories," JHEP 0910, 015 (2009),
[arXiv:0908.0363 [hep-th]].
\bibitem{lif90} S. A. Hartnoll, D. M. Hofman, D. Vegh,``Stellar spectroscopy: Fermions and holographic Lifshitz criticality," JHEP 1108 (2011) 096.
\bibitem{lif9} E. J. Brynjolfsson, U. H. Danielsson, L. Thorlacius and T. Zingg, ``Holographic Superconductors with Lifshitz Scaling,"
J. Phys. A 43: 065401, 2010,
[arXiv:0908.2611 [hep-th]].
\bibitem{DWP}D. W. Pang, ``On Charged Lifshitz Black Holes,"  JHEP {\bf 1001} (2010) 116,
[arXiv:0911.2777 [hep-th]].
\bibitem{DMP}M. H. Dehghani, R. Pourhasan, R. B. Mann, ``Charged Lifshitz Black Holes," Phys. Rev. D 84, 046002 (2011),
[arXiv:1102.0578 [hep-th]].
\bibitem{sto} U. Gursoy, E. Plauschinn, H. Stoof and S. Vandoren, ``Holography and ARPES sum-rules," [arXiv:1112.5074 [hep-th]].
\bibitem{AMM} M. Alishahiha, M. R. M. Mozaffar, A. Mollabashi, ``Fermions on Lifshitz Background," [arXiv:1201.1764 [hep-th]].
\bibitem{dipole} M. Edalati, R. Leigh, and P. W. Phillips, ``Dynamically generated Mott gap from holography,"
Phys. Rev. Lett. 106, 091602 (2011), [arXiv:1010.3238 [hep-th]].\\
M. Edalati, R. Leigh, K. W. Lo, and P. W. Phillips, ``Dynamical gap and Cuprate-like physics
from holography," Phys. Rev. D 83,(2011) 046012, [arXiv:1012.3751 [hep-th]].\\
W. J. Li, H. Zhang,``Holographic non-relativistic fermionic fixed point and bulk dipole coupling," JHEP 11 (2011) 018,
[arXiv:1110.4559 [hep-th]].\\
W. J. Li, R. Meyer, H. Zhang, ``Holographic non-relativistic fermionic fixed point by the charged dilatonic black hole," JHEP 01 (2012) 153,
[arXiv:1111.3783 [hep-th]].\\
J. P. Wu, H. B. Zeng, ``Dynamic gap from holographic fermions in charged dilaton black branes," JHEP 04 (2012) 068, 
[arXiv:1201.2485 [hep-th]].\\
W. Y. Wen, S. Y. Wu, ``Dipole Coupling Effect of Holographic Fermion in Charged Dilatonic Gravity,"
[arXiv:1202.6539 [hep-th]].\\
W. J. Li, J. P. Wu, ``Holographic fermions in charged dilaton black branes,"
[arXiv:1203.0674 [hep-th]].\\
X. M. Kuang, B. Wang, J. P. Wu,``Dipole Coupling Effect of Holographic Fermion in the Background of Charged Gauss-Bonnet AdS Black Hole,"
JHEP 07 (2012) 125, [arXiv:1205.6674[hep-th]].\\
X. M. Kuang, B. Wang, J. P. Wu,``Dynamical gap from holography in the charged dilaton black hole,"
[arXiv:1210.5735[hep-th]].\\
\bibitem{IL} N. Iqbal and H. Liu, ``Real-time response in AdS/CFT with application to spinors,¡±
Fortsch. Phys. 57, 367 (2009), [arXiv:0903.2596 [hep-th]].
\bibitem{JPW2}J. P. Wu, ``Some properties of the holographic fermions in an extremal charged dilatonic black hole,"
Phys. Rev. D 84,064008(2011),
[arXiv:1108.6134 [hep-th]].
\bibitem{Senthil1}T. Senthil, ``Critical fermi surfaces and non-fermi liquid metals," Phys. Rev. B 78, 035103 (2008),
[arXiv:0803.4009 [cond-mat.str-el]].
\bibitem{Senthil2}T. Senthil, ``Theory of a continuous Mott transition in two dimensions," Phys. Rev. B 78, 045109 (2008),
[arXiv:0804.1555 [cond-mat.str-el]].
\bibitem{LHLuttinger}N. Iqbal, H. Liu, ``Luttinger's theorem, superfluid vortices, and holography," Class. Quant. Grav. 29, 194004 (2012),
[arXiv:1112.3671 [hep-th]].
\bibitem{sch}S. Sachdev, ``A model of a Fermi liquid using gauge-gravity duality," Phys. Rev. D 84, 066009 (2011).
\bibitem{mc} A. Allais, J. McGreevy, S. J. Suh,``A quantum electron star," Phys. Rev. Lett. 108, 231602 (2012).
\end{thebibliography}
\end{document}